\documentclass[aps,prc,twocolumn,superscriptaddress,nofootinbib,showpacs,showkeys,preprintnumbers]{revtex4-1}
\usepackage{graphicx}  
\usepackage{bm}        
\usepackage{amssymb}   
\usepackage{amsmath}   
\usepackage{hyperref}
\usepackage{lineno}

\usepackage[usenames,dvipsnames]{xcolor}

\newcommand\numberthis{\addtocounter{equation}{1}\tag{\theequation}}

\def\snn{\mbox{$\sqrt{s_{_{\rm NN}}}$}}

\newcommand{\trento}{T\raisebox{-2pt}{R}ENTo}

\begin{document}
\title{Bulk pressure in fluid-dynamical simulations of Pb-Pb and p-Pb collisions at the LHC energies}
\author{Josef Bobek} \affiliation{Faculty of Nuclear Sciences and Physical Engineering, Czech Technical University in Prague,\\  B\v rehov\'a 7, 11519 Prague 1, Czech Republic}
\author{Iurii Karpenko} \affiliation{Faculty of Nuclear Sciences and Physical Engineering, Czech Technical University in Prague,\\  B\v rehov\'a 7, 11519 Prague 1, Czech Republic}

\begin{abstract}
State-of-the-art fluid dynamical simulations of relativistic heavy-ion collisions employ initial state models which result in a rather strong radial flow. In order to fit the experimental observables, a non-negligible bulk viscosity of the QGP and/or hadronic matter is required. We examine modern parametrizations of the bulk viscosity to entropy density ratio $\zeta/s$ used in recent fluid dynamical simulations, and explore the relative magnitude of the associated bulk viscous corrections in space-time, for Pb-Pb and p-Pb collisions at $\snn=2.76$~TeV LHC energy with a state-of-the-art initial state provided by \trento\ model. It turns out that, at a typical particlization hypersurface, the effect of bulk viscosity out-competes the one of shear viscosity, and in a significantly large part of the space-time volume, the  bulk pressure strongly counteracts the equilibrium pressure, thus the effective pressure approaches zero. The latter potentially challenges the applicability of fluid-dynamical modelling of heavy ion-ion and proton-ion collisions at the LHC energies.
\end{abstract}
\maketitle

\section{Introduction}
Soon after the collective behaviour of matter in the state of Quark-Gluon Plasma (QGP), created in heavy-ion collisions at $\snn=200$~GeV was established, the goal has become to constrain the transport properties of the created medium. At present, there is a consensus in the field that the effective ratio of shear viscosity to entropy density of the QGP medium at the temperatures reached at the RHIC collider is a factor 1.0--2.0 times the minimal shear visosity $\eta/s=1/4\pi$, with slightly larger values corresponding to hotter fireballs created at the LHC.
However, the ratio of the bulk viscosity of QGP to its entropy density, $\zeta/s$, is currently less constrained than its shear counterpart.

Early viscous hydrodynamic simulations for Au-Au collisions at $\sqrt{s}=200$~GeV RHIC energy assumed rather small values of temperature-dependent bulk viscosity. For example, a minimally constructed temperature-dependent $[\zeta/s](T)$ in simulations by Song and Heinz \cite{Song:2009gc} peaks around $T=175$~MeV with a value of about 0.04, while in their later calculations \cite{Song:2010aq} the bulk viscosity was omitted. Hydrodynamic simulations by Piotr Bozek \cite{Bozek:2011ua} employed $[\zeta/s](T)$ which is non-zero only in the hadronic phase and around the phase transition and has the same peak value of 0.04. Such $[\zeta/s](T)$ resulted in a rather small bulk pressure, with most of the bulk viscosity effect coming from the increase of $[\zeta/s](T)$ towards the phase transition and in the hadronic phase. The bulk viscous corrections had a small if not negligible influence on the evolution of the dense QGP medium created in relativistic heavy-ion collisions.

Recently, a new generation of fluid dynamical calculations has been introduced \cite{Ryu:2015vwa, Denicol:2015nhu, Bernhard:2016tnd, Ryu:2017qzn, Schenke:2019ruo}. Both \cite{Ryu:2017qzn} and \cite{Schenke:2019ruo} employ a much improved initial state from the IP-Glasma framework \cite{Schenke:2012wb}, while \cite{Bernhard:2016tnd} employs \trento\ initial state model. In \cite{Denicol:2015nhu} an extended Monte Carlo Glauber initial state is used, with sub-nucleonic degrees of freedom included. All of the above-mentioned calculations agree that the bulk viscosity is essential for a consistent description of hadron multiplicities and average transverse momentum of hadrons in heavy-ion collisions with the state-of-the-art initial state models. In particular, \cite{Ryu:2015vwa} shows that the average transverse momentum $p_T$ of the identified hadrons is too large when only the shear viscosity is included in the simulation, while the bulk viscosity suppresses the transverse expansion, thereby decreasing the average $p_T$ down to the experimental values.

\begin{figure}
    \centering
    \includegraphics[width=0.5\textwidth]{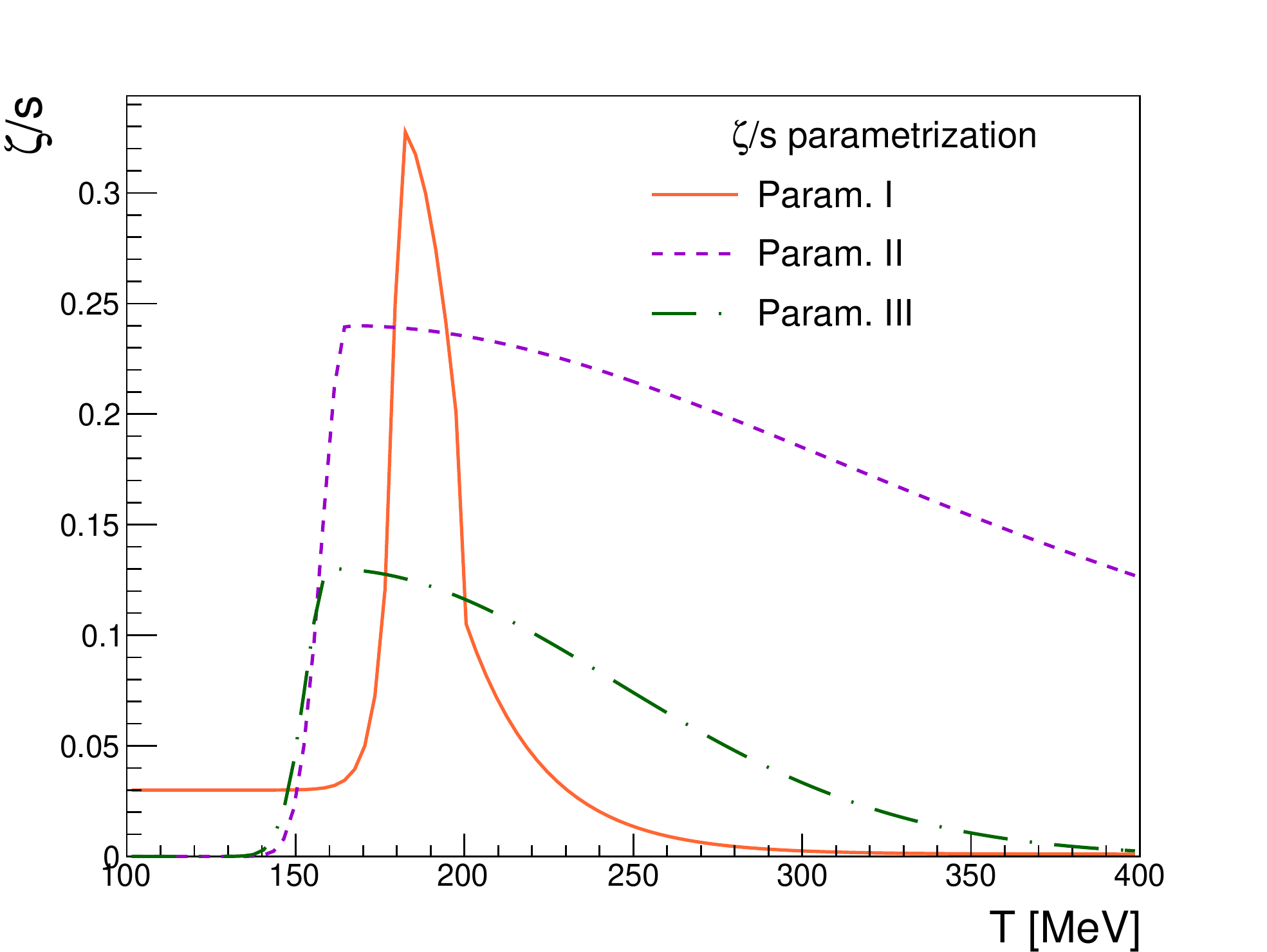}
    \caption{Modern $\zeta/s$ parametrizations used in literature: ``Param.~I'' is taken from \cite{Ryu:2017qzn}, ``Param.~II'' is taken from \cite{Schenke:2019ruo}, ``Param.~III'' from \cite{Schenke:2020mbo}. }
    \label{fig:zetas-parametrizations}
\end{figure}

Note that the calculations \cite{Ryu:2015vwa, Denicol:2015nhu, Bernhard:2016tnd, Ryu:2017qzn, Schenke:2019ruo} all assume $[\zeta/s](T)$ which peaks at much larger values, compared to the early studies. The modern parametrizations of $[\zeta/s](T)$ are conveniently summarized in \cite{Byres:2019xld} and denoted therein as ``Param.\ I'', ``Param.\ II'', which are also shown in Fig.~\ref{fig:zetas-parametrizations}. Note that in Param.\ I $[\zeta/s](T)$ has a rather narrow peak with a peak value of around 0.3, whereas Param.\ II has a broad peak with a maximum value of 0.24. Furthermore, \cite{Byres:2019xld} concludes that $[\zeta/s](T)$ parametrizations I and II result in a bulk pressure $\Pi$ which overturns the equilibrium pressure $p$, so that the effective pressure $\Pi+p$ becomes negative. This conclusion is, of course, initial-state dependent, and indeed, \cite{Byres:2019xld} shows that the regions of negative effective pressure emerge when the initial energy density profile for the hydrodynamic simulation is a Gaussian with a width of about 1 fm. Such an initial state is rather appropriate to assess high-multiplicity $pp$ collisions with a hydrodynamic approach, and it is known that the Gaussian initial state profile facilitates a rapid development of transverse expansion in the bulk. The study \cite{Byres:2019xld} did not proceed to assess the development of the bulk pressure in a scenario with a realistic initial state corresponding to heavy-ion collisions.

Later, in \cite{Cheng:2021tnq} it was demonstrated that both the necessary and the sufficient causality conditions for restricted and full Denicol-Niemi-Moln\'{a}r-Rischke (DNMR) theories of relativistic viscous hydrodynamics involve the ratio of the bulk pressure to the enthalpy density $\Pi/(\epsilon+p)$, which is the inverse Reynolds number associated with the bulk viscosity.  Another study of causality conditions in fluid dynamical simulations of Pb+Pb collisions at an LHC energy with initial conditions from \trento\ and \trento+Kompost has been carried out in \cite{Plumberg:2021bme}. Both studies employ yet another parametrization of $[\zeta/s](T)$, taken from \cite{Schenke:2020mbo}, which we dub as Param.~III, and which is approximately twice smaller than the Param.~II. The latter study finds that most of the violations of necessary and/or sufficient causality conditions occur at early stages of evolution or at the periphery of the fireball. However, our intuition tells us that, while the shear stress tensor generally tends to grow at the periphery of the expanding fireball, the same may not be true for the bulk pressure.

Therefore, in this study we follow the spirit of \cite{Byres:2019xld} and access whether a significant negative bulk pressure develops during the hydrodynamic simulation of Pb-Pb and $p$-Pb collisions at the LHC energies with the state-of-the-art initial state from the \trento model and with state-of-the-art $[\zeta/s](T)$ parametrizations.

\section{Model}
We simulate the expansion of hot and dense matter in Pb-Pb collisions at $\snn=2760$~GeV with a 3+1 dimensional relativistic viscous hydrodynamic code \texttt{vHLLE} \cite{Karpenko:2013wva}. The initial conditions are computed with the reduced thickness ansatz $T_R=\sqrt{T_A T_B}$ using the \trento code \cite{Moreland:2014oya}. The $T_R$ defines the initial entropy density profile in the transverse plane. On top of that, a finite longitudinal profile of the initial state is assumed, the same as in \cite{Cimerman:2020iny}.

The code solves relativistic viscous hydrodynamic equations in the following form:
\begin{equation}
 \partial_\nu T^{\mu\nu}=0, \quad T^{\mu\nu}=(\epsilon+p)u^\mu u^\nu - (p+\Pi)g^{\mu\nu}+\pi^{\mu\nu}
\end{equation}
\begin{align*}
    D\Pi=&\frac{-\zeta\theta-\Pi}{\tau_\Pi}-\frac{\delta_{\Pi\Pi}}{\tau_\Pi}\Pi\theta+\frac{\lambda_{\Pi\pi}}{\tau_\Pi}\pi^{\mu\nu}\sigma_{\mu\nu},\numberthis \label{equation:IS1}
    \\
    D\pi^{\langle\mu\nu\rangle} =& \frac{2\eta\sigma^{\mu\nu}-\pi^{\mu\nu}}{\tau_\pi}-
   \frac{\delta_{\pi\pi}}{\tau_\pi}\pi^{\mu\nu}\theta + \frac{\phi_7}{\tau_\pi}\pi_\alpha^{\langle\mu}\pi^{\nu\rangle\alpha}
   \\
   &-\frac{\tau_{\pi\pi}}{\tau_\pi}\pi_\alpha^{\langle\mu}\sigma^{\nu\rangle\alpha} + \frac{\lambda_{\pi\Pi}}{\tau_\pi}\Pi\sigma^{\mu\nu}.\numberthis \label{equation:IS2}
\end{align*}
Unlike all previous studies with the \texttt{vHLLE} code, the bulk viscosity is not set to zero. More specifically, we employ 3 different parametrizations of temperature-dependent $\zeta/s$, which we have briefly discussed in the Introduction, show in Fig.~\ref{fig:zetas-parametrizations} and label them as ``Param.~I'', ``Param.~II'' and ``Param.~III''. Parametrization I was introduced in \cite{Ryu:2017qzn} and has the following functional form:
\begin{equation}\label{equation:param1}
\zeta / s=\begin{cases}
c_1+0.08 \exp\left[\frac{T/T_{p}-1}{0.0025}\right]+
&\\ \quad+0.22 \exp\left[\frac{T / T_{p}-1}{0.022}\right]
& T<T_p
\\
c_2+27.55\left(T / T_{p}\right)-
&\\ \quad-13.77\left(T / T_{p}\right)^{2} 
& T_p<T<T_P
\\
c_3+0.9 \exp\left[\frac{-\left(T / T_{p}-1\right)}{0.0025}\right]+
&\\ \quad+0.25 \exp\left[\frac{-\left(T / T_{p}-1\right)}{0.13}\right] 
& T>T_P,
\end{cases}
\end{equation}
where $T_p=180$~MeV, $T_P=200$~MeV, $c_1=0.03$, $c_2=-13.45$ and $c_3=0.001$. A more recent $[\zeta/s](T)$ parametrization, introduced in \cite{Schenke:2019ruo} has a large $\zeta/s$ over a rather broad temperature range. We label it as Parametrization II:
\begin{equation}\label{equation:param2}
\zeta / s=\begin{cases}
B_{\mathrm{norm}}\exp\left[-\left(\frac{T-T_{\mathrm{peak}}}{T_{\mathrm{width}}}\right)^2\right]
& T<T_{\mathrm{peak}}
\\
B_{\mathrm{norm}}\frac{B_{\mathrm{width}}^2}{(T/T_{\mathrm{peak}}-1)^2+B_{\mathrm{width}}^2}
& T>T_{\mathrm{peak}},
\end{cases}
\end{equation}
where $T_{\mathrm{peak}}=165$~MeV, $B_{\mathrm{norm}}=0.24$, $B_{\mathrm{width}}=1.5$ and $T_{\mathrm{width}}=50$~MeV. The last parametrization (Parametrization III) employed is also the most recent one, introduced in \cite{Schenke:2020mbo}. It has a smaller peak value and also a more narrow peak, compared to Param.~II:
\begin{equation}\label{equation:param3}
\zeta / s=\begin{cases}
B_{\rm n}\exp\left[-\left(\frac{T-T_{\mu}}{B_1}\right)^2\right]
& T<T_{\mu}
\\
B_{\rm n}\exp\left[-\left(\frac{T-T_{\mu}}{B_2}\right)^2\right]
& T>T_{\mu},
\end{cases}
\end{equation}
where $B_{\rm n}=0.13$, $B_1=10$~MeV, $B_2=120$~MeV and $T_{\mu}=160$~MeV.

The hydrodynamic evolution starts with zero viscous corrections, which represents a lucky but unlikely scenario when the large early-time viscous corrections mentioned in \cite{Plumberg:2021bme} are not an issue.

\begin{figure}
    \centering
    \includegraphics[width=0.49\textwidth]{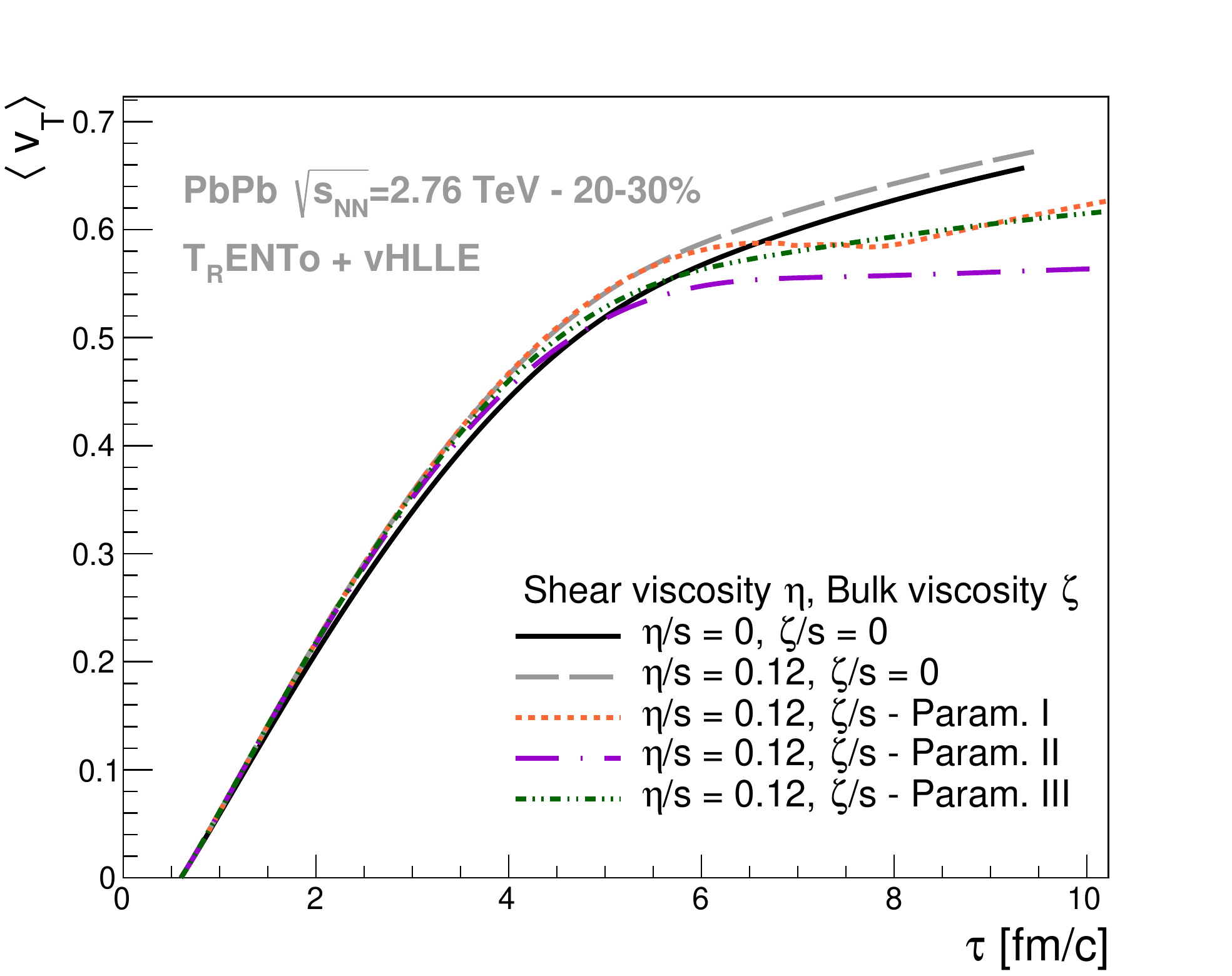}
    \caption{Time dependence of radial flow $\langle v_T\rangle$ in ideal and viscous fluid dynamic simulations of 20-30\% central Pb-Pb collisions at $\snn=2.76$~TeV.}
    \label{fig:radFlow}
\end{figure}

\section{Results and discussion}
With the setup from the previous Section, we simulate 0-5\%, 20-30\% and 40-50\% central Pb-Pb collisions at $\snn=2.76$~TeV, as well as $p$-Pb collisions at $\snn=5.02$~TeV. First in Fig.~\ref{fig:radFlow} we plot the time evolution of radial flow for 20-30\% centrality in the cases of ideal fluid ($\eta/s=\zeta/s=0$), and viscous fluid with constant shear viscosity ($\eta/s=0.12$) and four different scenarios for $\zeta/s$: zero, Param.~I, Param.~II, Param.~III. To avoid overfilling the text with plots, we keep the rest of the cases (Param.~I and Param.~III, and 0-5\%, 40-50\% central Pb-Pb) in the supplementary material.

In Fig.~\ref{fig:radFlow} we observe familiar effects: shear viscosity somewhat enhances the radial flow, by redistributing the energy from the strong and pre-dominant longitudinal expansion into the transverse expansion. Such redistribution starts to happen early on, therefore the slope of $v_T(\tau)$ is slightly larger for the $\eta/s=0.12$ case as compared to the $\eta/s=0$ case from the very beginning. On the contrary, the bulk viscosity does not influence the radial flow of the bulk at first; however, it starts to slow down the pace of expansion at late times.

Bulk viscosity affects the geometry and transverse flow at a typical particlization/freeze-out much stronger than the shear viscosity. This can be seen in Fig.~\ref{fig:freezeout}, where the transverse flow velocity profiles at an isotherm of $T_c=165$~MeV are plotted as a function of the Bjorken time $\tau$ (on the Y axis). The effect in the particlization hypersurface is stronger because $[\zeta/s](T)$ peaks around temperatures close to $T_c$. Modification of transverse flow velocity at the particlization echoes in Fig.~\ref{fig:radFlow} at late times, when the fluid stage ends for the bulk medium at around $\tau=9$~fm/c.

\begin{figure*}[htb]
    \vspace{-20pt}
    \centering
    \includegraphics[width=0.33\textwidth]{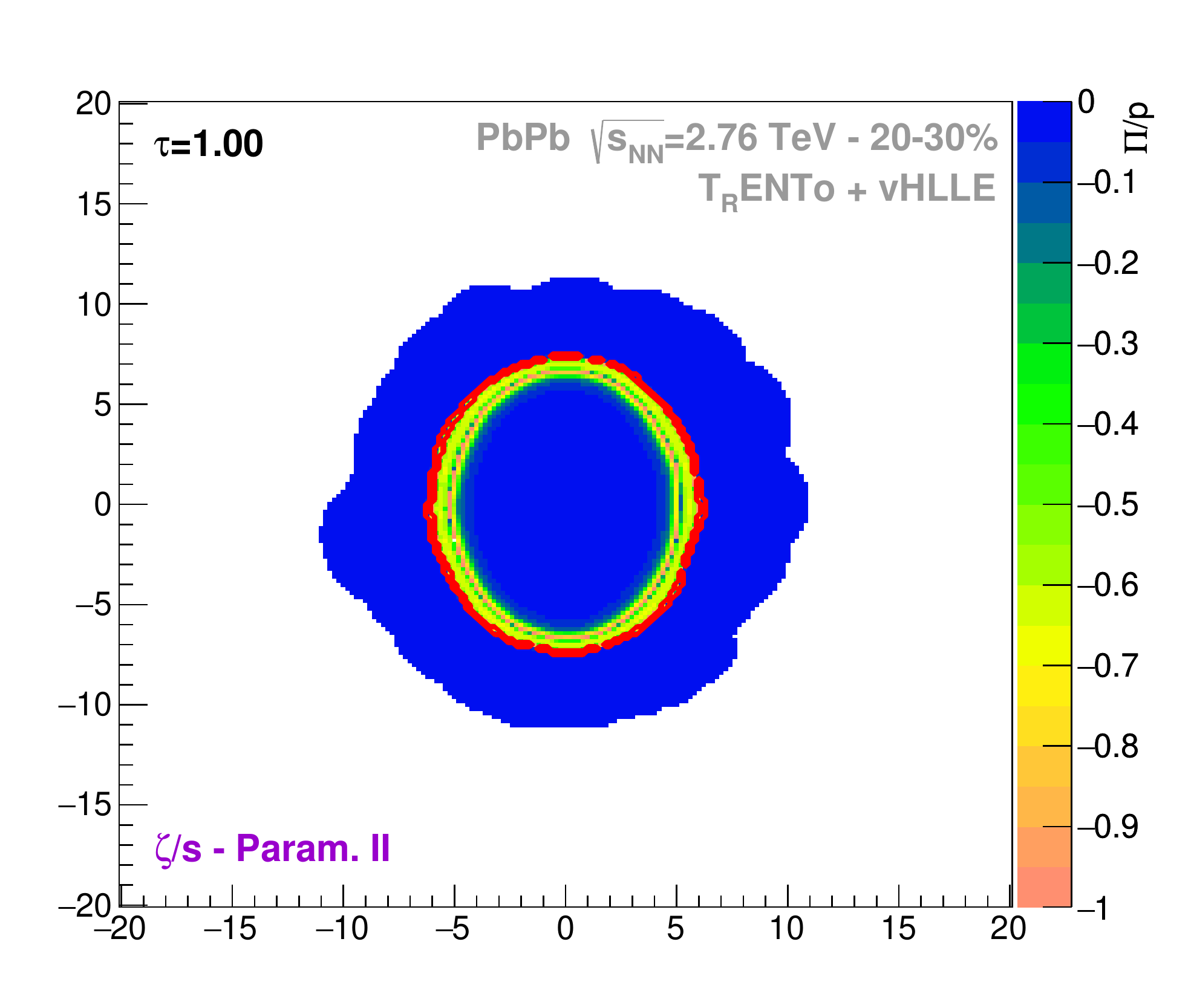}\hspace{-5pt}
    \includegraphics[width=0.33\textwidth]{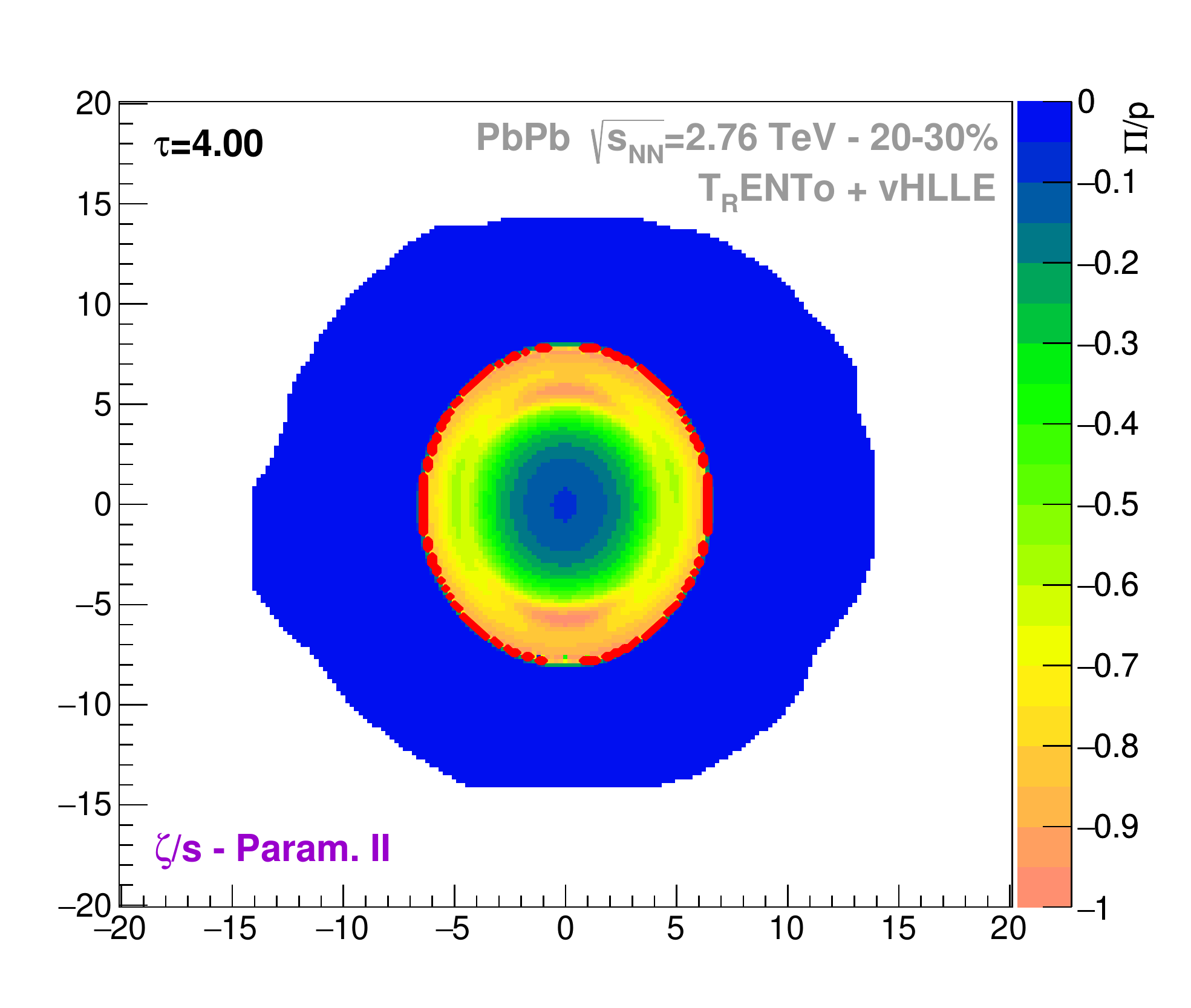}\hspace{-5pt}
    \includegraphics[width=0.33\textwidth]{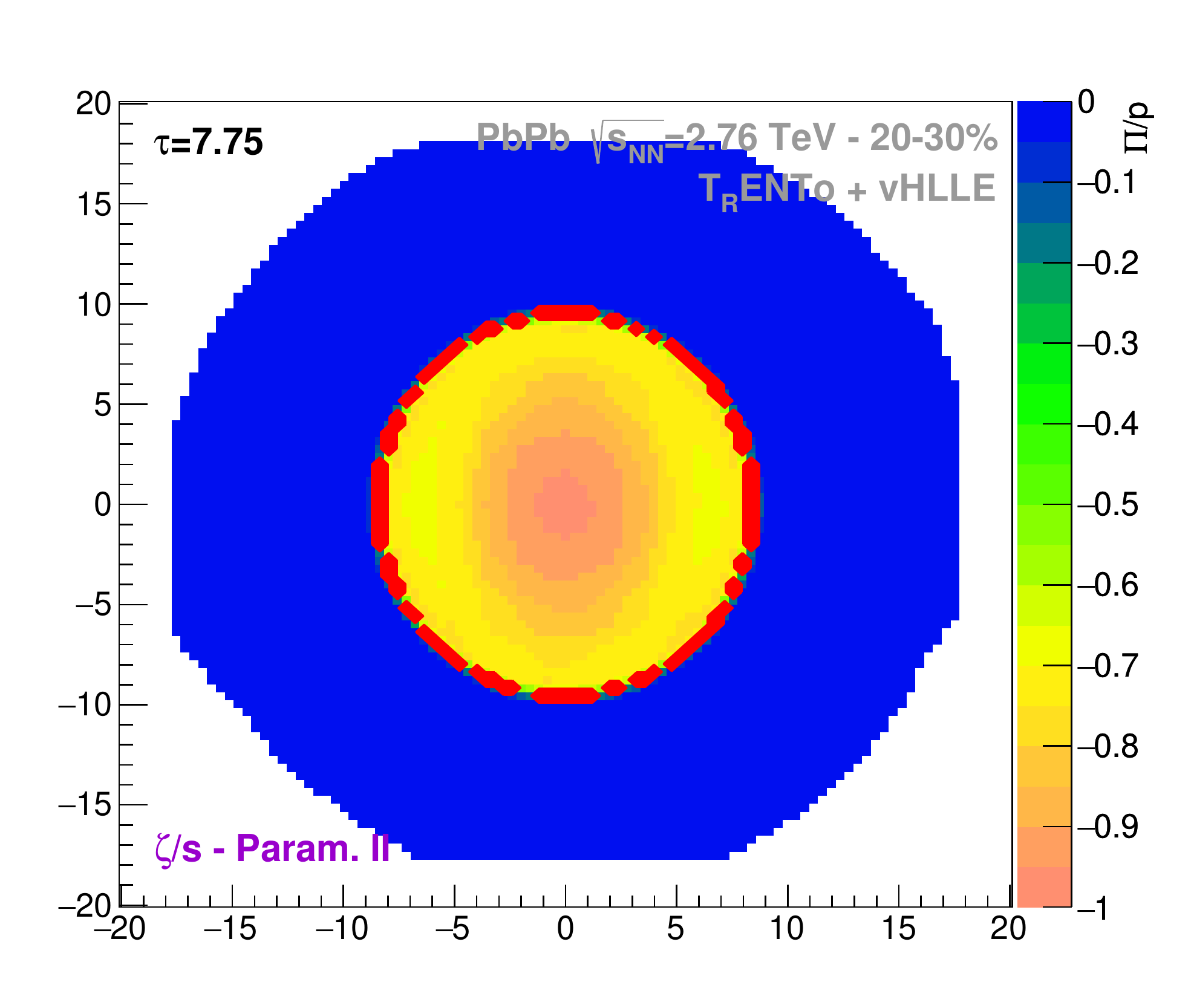}\\
    \includegraphics[width=0.33\textwidth]{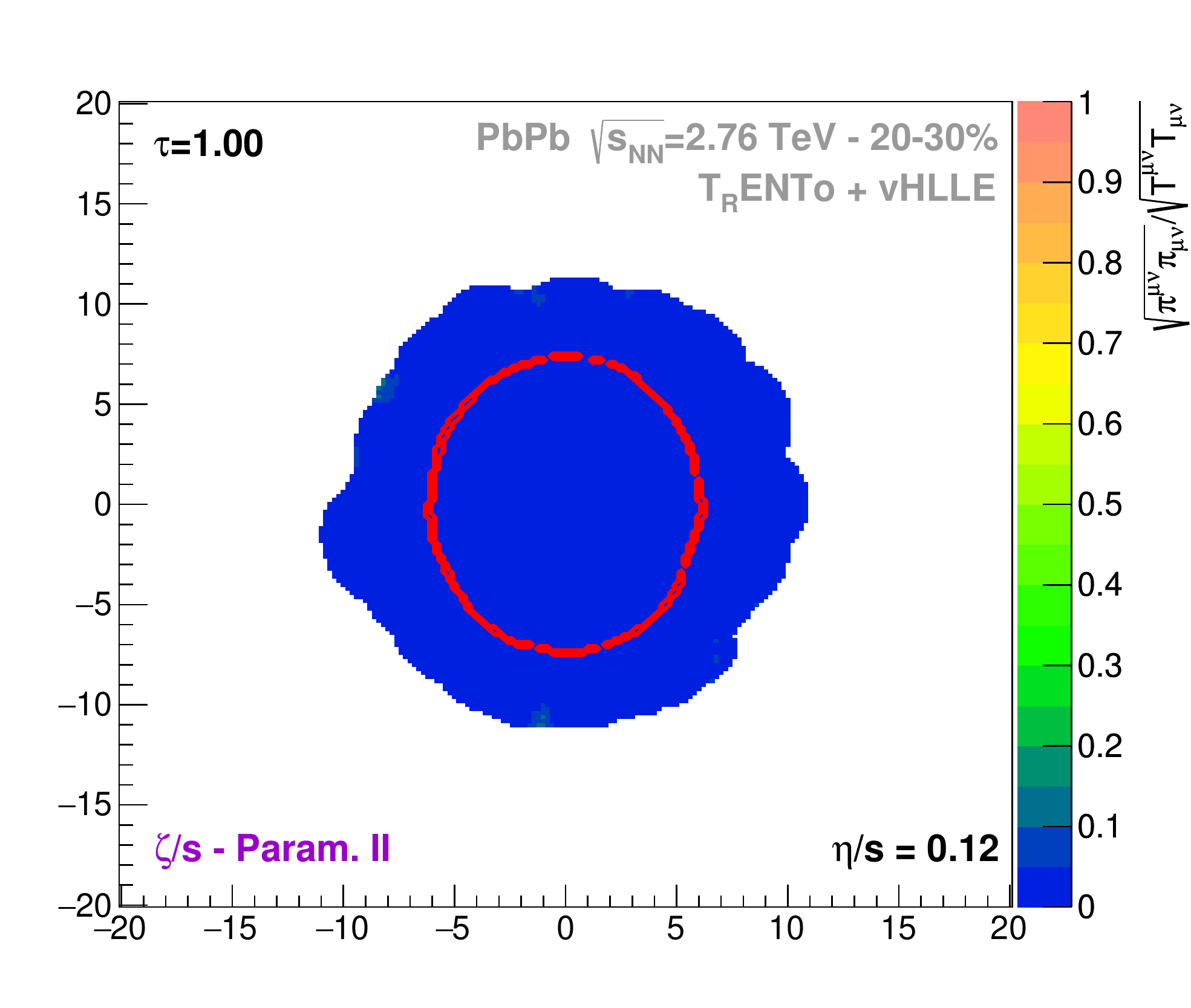}\hspace{-5pt}
    \includegraphics[width=0.33\textwidth]{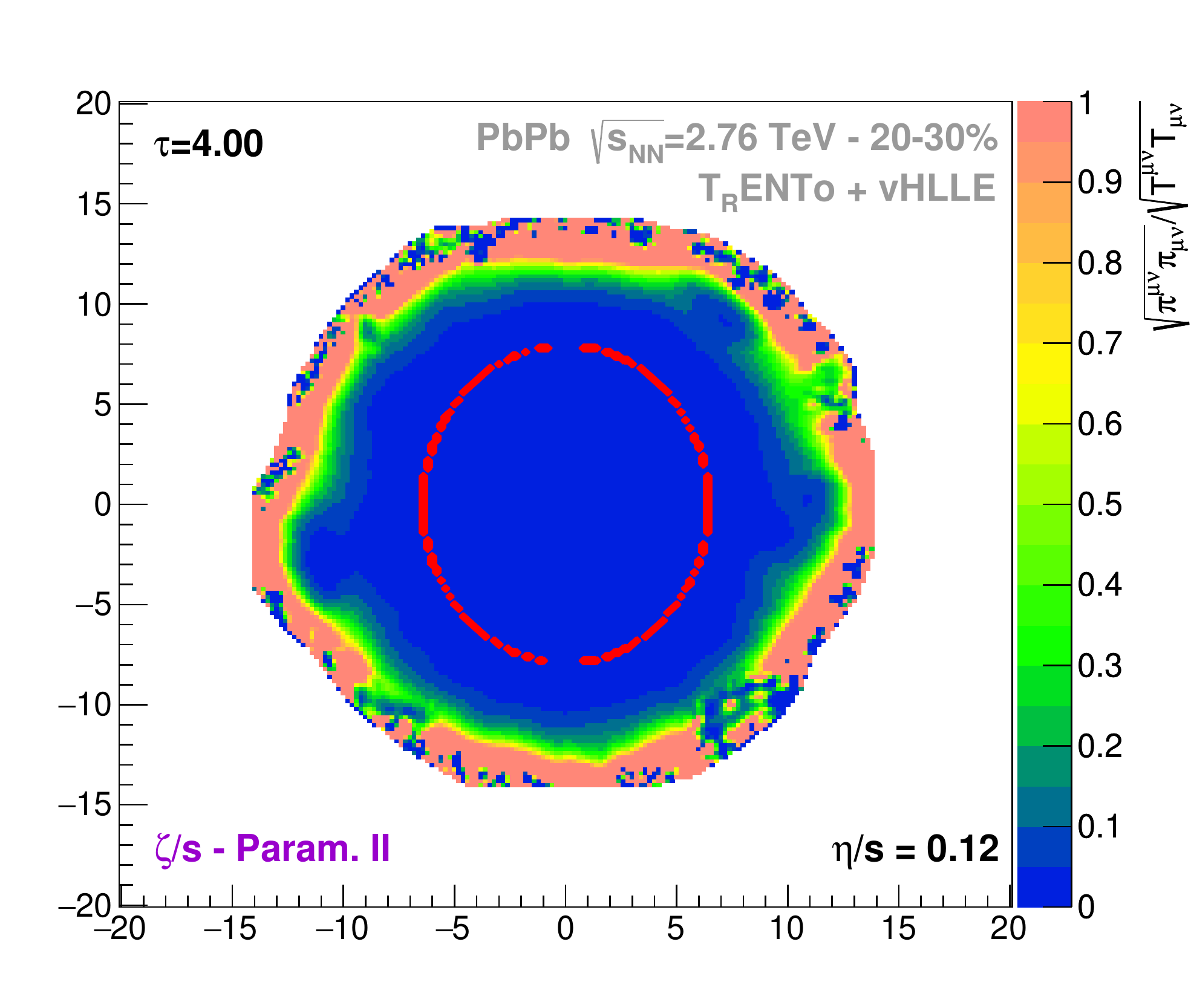}\hspace{-5pt}
    \includegraphics[width=0.33\textwidth]{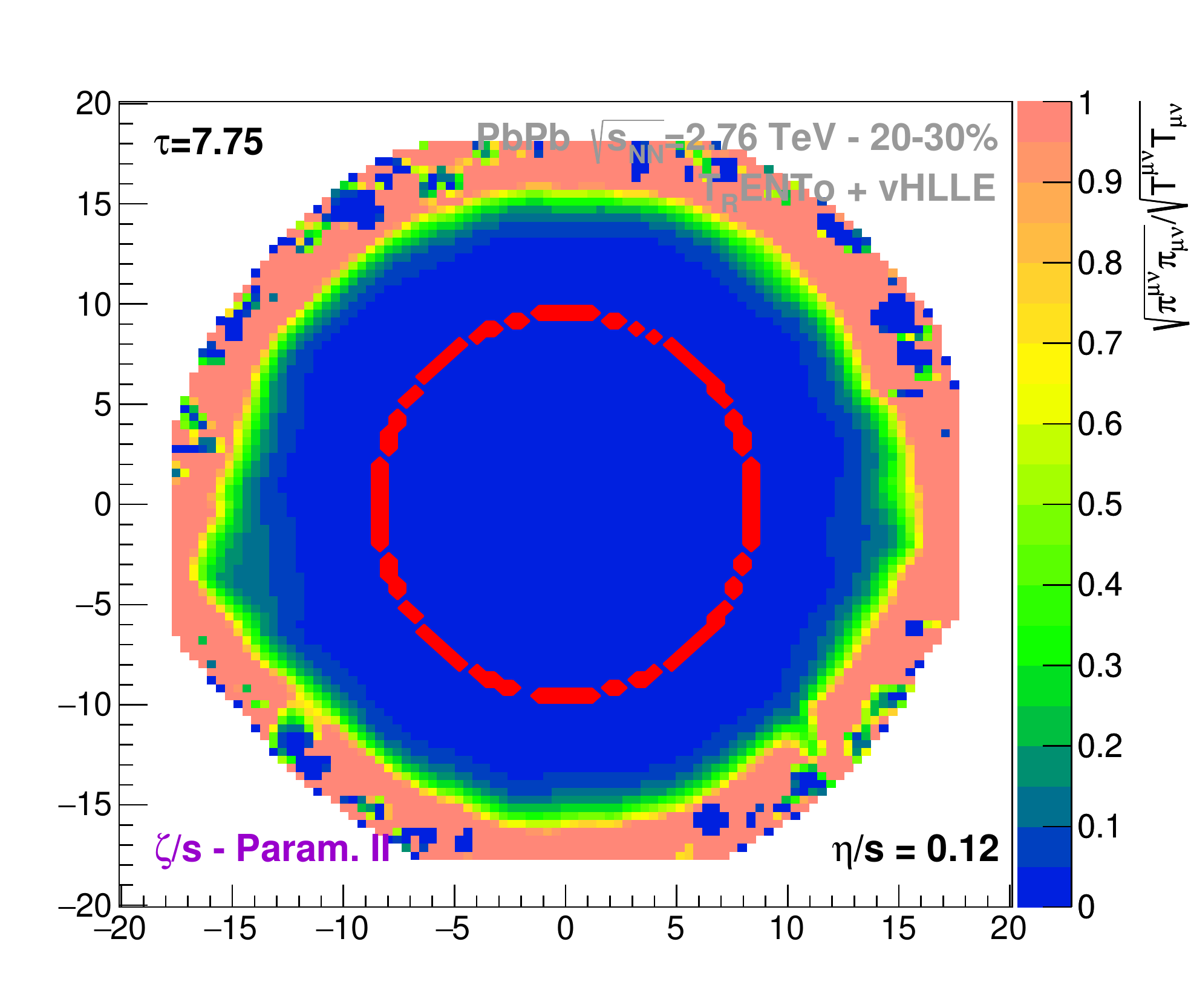}
    \caption{$\Pi/p$ (top panel) and $\sqrt{\pi^{\mu\nu}\pi_{\mu\nu}}/\sqrt{T_0^{\mu\nu}T_{0,\mu\nu}}$ (bottom panel) ratios at zero space-time rapidity $\eta_s$ and different times in hydrodynamic simulation of 20-30\% central Pb-Pb collisions at $\snn=2.76$~TeV with \trento\ initial state.}
    \label{fig:Pip}
\end{figure*}

How large does the bulk pressure actually get during the hydrodynamic evolution? To answer that, we visualised the distribution of the ratio of bulk pressure $\Pi$ to equilibrium pressure $p$ in hydrodynamic cells at zero space-time rapidity $\eta_s=0$ at different times $\tau$ in the top panel of Fig.~\ref{fig:Pip}. For this visualization, $\zeta/s$ Param.~II was used. One can see that, whereas at early times the bulk pressure is comparatively small, its magnitude starts to grow towards late times, approaching the magnitude of the equilibrium pressure. The bulk pressure does not grow further simply because the hydrodynamic code does not allow it to. Technically it limits the bulk pressure so that its magnitude is not larger than that of the equilibrium pressure. This growth of bulk pressure manifests itself again in Fig.~\ref{fig:radFlow}, where the curve corresponding to Param.~II flattens after $\tau=6$~fm/c, meaning that the transverse expansion does not accelerate any longer despite still significant energy density and equilibrium pressure gradients in the system. In addition, Fig.~\ref{fig:Pip} shows the sections of iso-therm $T=150$~MeV at given $\tau$ by the red line. One can see that the regions of large bulk pressure stop at around the same temperature $T=150$~MeV, mainly because with Parametrisation II, $\zeta/s$ goes to zero at lower temperatures.
\begin{figure*}[htb]
    \centering
    \includegraphics[width=0.9\textwidth]{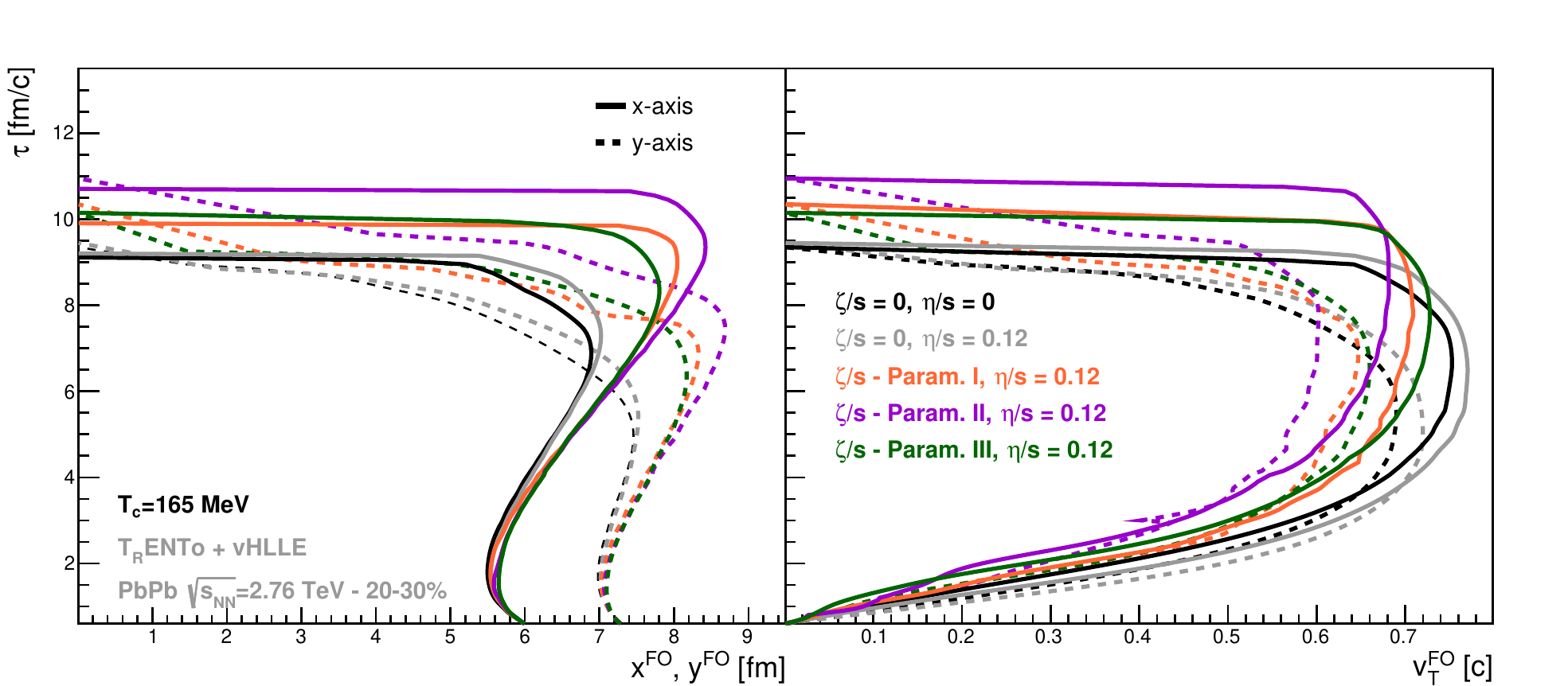}
    \caption{Relation of proper time $\tau$ and transverse size of freeze-out hyper-surface, $x^{\mathrm{FO}}$ and $y^{\mathrm{FO}}$ (left panel), and transverse velocity at freeze-out hyper-surface, $v_T^{\mathrm{FO}}$ (right panel) at zero space-time rapidity $\eta_s=0$ in ideal and viscous fluid dynamics of 20-30\% central Pb+Pb \texttt{vHLLE} 3-dimensional simulations at $\sqrt{s_{NN}}=2.76$~TeV using \trento\ initial conditions.}
    \label{fig:freezeout}
\end{figure*}

For comparison, we plot a similar measure of the relative magnitude of the shear stress tensor, which is a ratio $\sqrt{\pi^{\mu\nu}\pi_{\mu\nu}}/\sqrt{T_0^{\mu\nu}T_{0,\mu\nu}}$ in the bottom panel of Fig.~\ref{fig:Pip}. Here, $T^{\mu\nu}_0$ is the ideal part of the energy-momentum tensor of the fluid. One can see that, unlike the bulk pressure, the shear stress tensor remains a relatively small correction in the core of the system. The $\pi^{\mu\nu}$ grows to large values only at the periphery of the fireball, where the energy density and pressure are small, and the 4-velocity gradients are large due to large gamma factor. It is precisely the periphery of the fireball where $\pi^{\mu\nu}$ limiters are typically engaged in hydrodynamic codes, in order to keep the evolution stable.

\begin{figure*}[htb]
    \centering
    \includegraphics[width=0.9\textwidth]{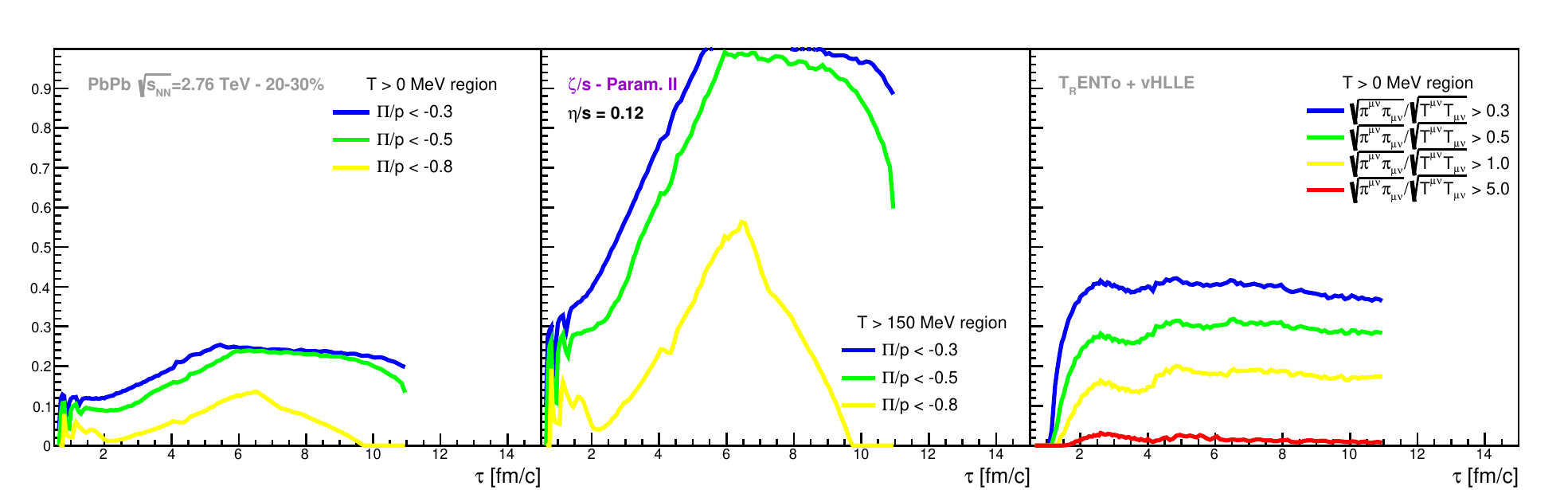}\\
    \includegraphics[width=0.9\textwidth]{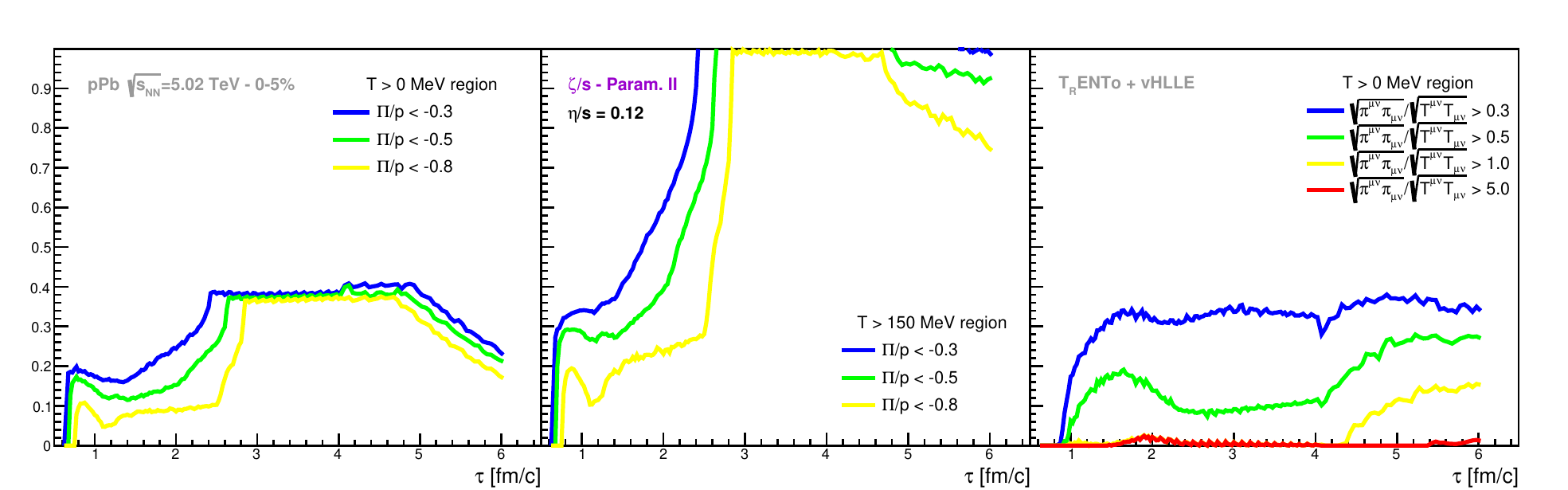}
    \caption{Fractions of fluid cells where the magnitude of bulk pressure $|\Pi|$ is larger than 0.3, 0.5 and 0.8 of equilibrium pressure $p$ (left and middle panels) and percentage of fluid cells where $\sqrt{\pi^{\mu\nu}\pi_{\mu\nu}}/\sqrt{T_0^{\mu\nu}T_{0,\mu\nu}}$ is larger than 0.3, 0.5, 1.0 and 5.0 (right panel). The middle panel correspond to $T>150$~MeV region only, whereas the left and the right panels correspond to all fluid cells, at zero space-time rapidity in all cases. The top panel corresponds to fluid dynamical simulations of 20-30\% central Pb-Pb collisions at $\snn=2.76$~TeV, whereas the bottom panel corresponds to $p$-Pb collisions at $\snn=5.02$~TeV}
    \label{fig:acausal}
\end{figure*}

The relative magnitude of bulk pressure is presented in a more condensed form in Fig.~\ref{fig:acausal}, where the time dependence of the fractions of fluid cells where the bulk pressure is less than -0.3, -0.5 and -0.8 times the equilibrium pressure, and where the $\sqrt{\pi^{\mu\nu}\pi_{\mu\nu}}$ is larger than 0.3, 0.5, 1.0, and 5.0 times $\sqrt{T_0^{\mu\nu}T_{0,\mu\nu}}$. The more condensed form allows to plot the time dependencies for both 20-30\% central Pb-Pb and most central p-Pb collision systems, and therefore to explore the system size dependence. We observe that for the 20-30\% Pb-Pb case, the fraction of fluid cells where the bulk pressure is at least -0.5 of the equilibrium pressure, raises with the evolution time $\tau$ rather quickly and stays close to 100\% throughout the remaining hydrodynamic expansion, up until the moment when the matter is in hadronic phase everywhere. The fraction of fluid cells where $\Pi<-0.8p$, peaks at 50\% around $\tau=6$ fm/c and then drops to nearly zero. In the case of p-Pb, due to the smaller initial transverse size and more violent expansion, all fractions saturate faster, and the fraction of cells where $\Pi<-0.8p$ is higher and does not drop with evolution time. This indicates that the relatively large bulk pressure developed with $[\zeta/s](T)$ Param.~II is certainly more of an issue for the fluid dynamical modelling of matter expansion as a whole, for the p-Pb system compared to a mid-central Pb-Pb system.

We summarise the findings so far as follows. For a given $[\zeta/s](T)$, all the fractions grow from 0-5\% central towards 40-50\% central Pb-Pb collisions. This trend is expected as the initial system size decreases from central to peripheral collisions, therefore leading to a stronger transverse expansion. $[\zeta/s](T)$ Param.~III is much more comfortable for hydrodynamic modelling, where in most scenarios the fraction of cells with $\Pi<-0.8p$ is small at all times. With $[\zeta/s](T)$ Param.~I, the fractions are also somewhat smaller compared to Param.~II. Therefore, unsurprisingly $[\zeta/s](T)$ Param.~II represents the most challenging scenario for fluid-dynamic modelling.

Once the bulk pressure grows large enough in magnitude, it counteracts the equilibrium pressure, so that the effective pressure $\Pi+p$ becomes small or even negative. A corresponding phenomenon in classical physics is called cavitation, when bubbles of gas start to form inside the fluid. In the case of heavy-ion collisions and QGP fluid, bubbles of hadron gas should start to form once the effective pressure becomes small. However, to our knowledge, none of the fluid dynamical codes for heavy-ion simulations is capable of treating such regime of fluid expansion.

As recent studies \cite{Cheng:2021tnq,Plumberg:2021bme} have shown, local bulk pressure enters into causality constraints for relativistic viscous fluid dynamics. More precisely, in the case of full DNMR theory, one of the necessary causality conditions which involves bulk viscosity is $|\Pi/(\epsilon+p)|<1$, whereas a corresponding sufficient condition is $|\Pi/(\epsilon+p)|\ll1$. The ratio $\Pi/(\epsilon+p)=R_\Pi$ corresponds to a Knudsen number associated with bulk viscosity. The study \cite{Cheng:2021tnq} further elaborates that by limiting the bulk Knudsen number by $|R_\Pi|<p/(\epsilon+p)$, which is the same as requiring $|\Pi|<p$ (no cavitation), the DNMR viscous hydrodynamic evolution stays within causal domain, and such restriction is only applied to a small percentage of fluid cells. An interesting consequence of the $|\Pi|<p$ restriction is that the average transverse momentum $\langle p_T\rangle$ of hadrons produced out of the fluid, an increase in such case by more than 25\% as can be seen in Fig.~9 in \cite{Cheng:2021tnq}. Furthermore, we stress that both studies employ Param.~III for $[\zeta/s](T)$ from \cite{Schenke:2020mbo}, which is twice smaller than the Param.~II and the hydrodynamic simulations in \cite{Plumberg:2021bme} correspond to Au-Au collisions at the highest RHIC energy, where the expansion is somewhat slower compared to Pb-Pb collisions at the LHC.

Lastly, our findings have implications for the Bayesian analysis applied to the fluid-dynamic modelling of heavy-ion collisions. Namely, in order to constrain the $\zeta/s$ of the medium from the rigorous model-to-data comparison, a Bayesian framework would need the hydrodynamic model output in a rather wide range of $\zeta/s$. Our concern is that at the upper limit of $\zeta/s$ the resulting bulk pressure may become so large that the cavitation regime will occupy a dominant part of the space-time domain of the dense matter evolution. Therefore, in such a corner of the parameter space of the model, its output may not be reliable, this jeopardising the ability of the Bayesian framework to properly extract an optimal value of $\zeta/s$ or an optimal $[\zeta/s](T)$ parametrization.

Overall, we find that modern $[\zeta/s](T)$ parametrizations are not always in the comfortable range, so that the results of fluid-dynamic modelling can be trusted even for a relatively ``large'' Pb-Pb system. Therefore, one has to be cautious when enabling the bulk viscosity in the fluid-dynamical simulations of heavy-ion collisions.

\section{Conclusions}
In this study we have implemented 3 modern parametrizations of the temperature-dependent bulk viscosity of the strongly interacting medium, and run hydrodynamic simulations of Pb-Pb as well as p-Pb collisions at 2.76 TeV LHC energy. We find that for all the parametrizations, bulk viscosity strongly influences the geometry and radial flow at the freeze-out, out-competing the effects of shear viscosity. Even more interestingly, in a significantly large part of the space-time volume, the magnitude of the bulk pressure becomes comparable to the one of equilibrium pressure. Thus, the effective pressure approaches zero. This does not impose an immediate problem for the stability of the hydrodynamic solution, as the necessary causality conditions include an inequality $|\Pi/(\epsilon+p)|<1$, which is still satisfied in the vast majority of fluid cells.\par
However, sufficient causality conditions include $|\Pi/(\epsilon+p)|\ll1$, which is less easy to achieve in actual hydrodynamic simulations. Therefore, the large bulk pressure does pose a challenge not only for early-time dynamics but also late-time dynamics of relativistic heavy-ion collisions.

\section{Acknowledgements}
JB acknowlegdes support by the project Centre of Advanced Applied Sciences, No.~CZ.02.1.01/0.0/0.0/16-019/0000778,  co-financed by the European Union. IK acknowledges support by the Ministry of Education, Youth and Sports of the Czech Republic under grant ``International Mobility of Researchers – MSCA IF IV at CTU in Prague'' No.\ CZ.02.2.69/0.0/0.0/20\_079/0017983.
\bibliographystyle{h-physrev}
\bibliography{main}
\end{document}